\documentstyle[12pt,a4]{article}

\newcommand{\nit}{\noindent}
\newcommand{\nl}{\newline}
\newcommand{\np}{\newpage}
\newcommand{\be}{\begin{equation}}
\newcommand{\ee}{\end{equation}}
\newcommand{\ba}{\begin{array}}
\newcommand{\ea}{\end{array}}
\newcommand{\dsp}{\displaystyle}
\newcommand{\ct}{\cite}
\newcommand{\bit}{\bibitem}
\newcommand{\ag}{\alpha}
\newcommand{\bg}{\beta}
\newcommand{\gam}{\gamma}
\newcommand{\del}{\delta}
\newcommand{\uag}{\underline{\alpha}}
\newcommand{\ubg}{\underline{\beta}}

\newcommand{\eps}{\epsilon}

\newcommand{\th}{\theta}
\newcommand{\kg}{\kappa}
\newcommand{\lb}{\lambda}
\newcommand{\sg}{\sigma}

\newcommand{\og}{\omega}
\newcommand{\Del}{\Delta}
\newcommand{\Fg}{\Phi} 
\newcommand{\Gam}{\Gamma}

\newcommand{\Sg}{\Sigma}
\newcommand{\lh}{\left(}
\newcommand{\rh}{\right)}
\newcommand{\pl}{\partial}
\newcommand{\cL}{{\cal L}}
\newcommand{\Kh}{K\"{a}hler} 

\newcommand{\bz}{\bar{z}} 
\newcommand{\bps}{\bar{\psi}}
\newcommand{\bF}{\bar{F}} 

\newcommand{\vs}[1]{\vspace{#1 em}} 

\newcommand{\Slashed}{\hspace{-1.3ex}/\hspace{.2ex}}

\begin{document}

\pagestyle{empty}

\begin{flushright}
NIKHEF/98-028 \\
%CONCEPT \\
%not for distribution
\end{flushright}

\begin{center}
{\Large {\bf Matter coupling and anomaly cancellation in 
supersymmetric $\sg$-models}} \\
\vspace{5ex}

{\large S.\ Groot Nibbelink and J.W.\ van Holten}\\
\vspace{3ex}

{\large NIKHEF, P.O.\ Box 41882}\\
\vspace{3ex}

{\large Amsterdam NL}
\vspace{5ex}

{August 24, 1998}
\end{center}
\vspace{15ex}

\nit
{\small 
{\bf Abstract}\nl
Generalized matter couplings to four-dimensional supersymmetric sigma 
models on general \Kh\, manifolds are presented, preserving all 
holomorphic symmetries. Our generalization allows assignment 
of arbitrary $U(1)$ charges to additional matter fermions, in all representations of (the holomorphic part of) the isometry group. 
This can be used to eliminate unwanted $\gam_5$ anomalies, in 
particular for the U(1) symmetry arising from the complex structure 
of the target space. A consistent gauging of this isometry group, 
or any of its subgroups, then becomes possible. When gauged in the 
presence of a chiral scalar multiplet, the U(1) symmetry is broken 
spontaneously, generating a mass for the U(1) vector multiplet via 
the supersymmetric Higgs effect. As an example we discuss the case 
of the homogeneous coset space $E_6/SO(10) \times U(1)$.   
}

\np

\pagestyle{plain}
\pagenumbering{arabic}

\nit 
Supersymmetric sigma-models in four space-time dimensions 
are formulated in terms of chiral superfields $\Fg^{\ag}$,  
the physical components of which are complex scalars $z^{\ag}$
and (left-handed) chiral spinors $\psi_L^{\ag}$. Because of 
the complex nature of the fields and the restrictions imposed 
by supersymmetry, the target space of the scalars is a \Kh\,  
manifold \ct{Zum,AF}. 

For models with rigid supersymmetry in flat space-time, the 
kinetic part of the lagrangean is given in terms of a real 
composite vector superfield $K(\bar{\Fg}, \Fg)$ by the following 
supersymmetric expression 
\be 
\cL_{\sg}\, =\, K(\bar{\Fg}, \Fg)|_D\, =\, 
 - g_{\ag \uag}\, \lh \pl^{\mu} \bz^{\uag} \pl_{\mu} 
 z^{\ag} + \bps_L^{\uag}\, D\Slashed \psi_L^{\ag} \rh\, 
 +\, \frac{1}{8}\, R_{\ag \uag \bg \ubg}\, \bps_L^{\uag}\, 
 \gam^{\mu} \psi_L^{\ag}\, \bps_L^{\ubg}\, \gam_{\mu} \psi_L^{\bg}. 
\label{1} 
\ee
\nit 
The real symmetric scalar function $K(\bz,z)$ is the local  \Kh\, 
potential from which the complex hermitean metric $g_{\ag \uag}$ 
is derived as its mixed second derivative 
\be 
g_{\ag \uag}\, =\, K_{, \ag \uag}.
\label{2.1}
\ee
\nit
The corresponding connection and curvature components are  
\be 
\Gam_{\bg \gam}^{\;\;\;\;\ag}\, =\, g^{\uag \ag} g_{\bg \uag, \gam}, 
\hspace{2em} 
R_{\ag \uag \bg \ubg}\, =\, g_{\uag \gam}\, 
 \Gam_{\ag\bg,\, \ubg}^{\;\;\;\;\gam} . 
\label{2}
\ee 
\nit 
The covariant derivative of the spinor field is formed with the 
holomorphic pull-back of the connection $\Gam_{\bg \gam}^{\;\;\;\;\ag}$ 
\be
D_{\mu} \psi_L^{\ag}\, =\, \pl_{\mu} \psi_L^{\ag}\, -\, 
 \pl_{\mu} z^{\bg}\, \Gam_{\bg \gam}^{\;\;\;\;\ag} \psi_L^{\gam}. 
\label{3} 
\ee 
\nit
The lagrangean $\cL_{\sg}$ is by construction invariant under a 
$U(1)$ symmetry multiplying the superfields $\Fg^{\ag}$, hence all 
its components $(z^{\ag}, \psi_L^{\ag})$, by a universal phase 
factor $e^{i \th}$. In geometrical language the symmetry can be  
represented in terms of a holomorphic Killing vector $R^{\ag}_{\th}(z)$
by the transformations 
\be 
\del_\th z^{\ag}\, =\, \th R^{\ag}_{\th}(z)\, =\, i \th q_{(\ag)} z^{\ag}, 
\hspace{2em}
\del_{\th} \psi_L^{\ag}\, =\, \th R_{\th, \bg}^{\ag}\, \psi_L^{\bg}\, 
 =\, i \th q_{(\ag)} \psi_L^{\ag}. 
\label{4} 
\ee 
\nit 
Here the $q_{(\ag)}$ represent the $U(1)$ charges of the fields.
Other symmetries may be present, depending on the properties 
of the \Kh\, manifold on which the scalars take their values. 
In particular, there may be a larger set of holomorphic 
Killing vectors $R^{\ag}_i(z)$ defining a Lie-algebra with structure 
constants $f_{ij}^{\;\;\;k}$: 
\be 
R_i^{\bg} R_{j , \bg}^{\ag}\, -\, R_j^{\bg} R_{i, \bg}^{\ag}\, 
  =\, f_{ij}^{\;\;\;k} R_k^{\ag}.
\label{6}
\ee 
\nit 
Then the lagrangean (\ref{1}) is invariant under the infinitesimal 
transformations generated by the derivation $\del = \th^i \del_i$:
\be 
\ba{ll}
\del z^{\ag}\, =\, \th^i R_i^{\ag}(z), & \del \bz^{\uag}\, =\, 
 \th^i \bar{R}_i^{\uag}(\bz), \\
 & \\
\del \psi_L^{\ag}\, =\, \th^i R_{i,\, \bg}^{\ag}(z)\, \psi^{\bg}, & 
\del \bps_L^{\uag}\, =\, \th^i \bar{R}_{i,\,   \ubg}^{\uag}(\bz)\, 
 \bps^{\ubg}. 
\ea 
\label{7}
\ee 
\nit 
In general such transformations can be non-linear, requiring the 
introduction of a dimensional parameter $f$, conventionally the 
inverse mass associated with the breaking of these symmetries. 
As is well-known \ct{BW}-\ct{jw2}, the holomorphic Killing vectors 
can be derived from a set of real potentials $M_i(z,\bz)$ such that 
\be
R_i^{\ag}(z)\, =\, -i g^{\uag \ag} M_{i,\, \uag}, \hspace{2em} 
\bar{R}^{\uag}_i\, =\, i g^{\uag \ag} M_{i,\, \ag}. 
\label{8}
\ee 
\nit
Adjusting the constants in $M_i$ they transform in the 
adjoint representation:
\be 
\del_i M_j\, =\, R_i^{\ag} M_{j,\, \ag}\, +\, \bar{R}_i^{\uag}
 M_{j,\, \uag}\, =\, - i \lh R_i^{\ag} \bar{R}_{j \ag} - 
 R_j^{\ag} \bar{R}_{i \ag} \rh\, =\, f_{ij}^{\;\;k} M_k. 
\label{9} 
\ee 
\nit 
For semi-simple Lie algebras this relation can be inverted \ct{Aoy}, 
using the contravariant components of the structure constants 
normalized to $f_{ij}^{\;\;k} f^{ij}_{\;\;\;l} = 2 \del^k_l$:
\be
M_i\, =\, - \frac{i}{2}\, f^{jk}_{\;\;\;i}\, \lh R^{\ag}_j \bar{R}_{k\ag}
 - R^{\ag}_k \bar{R}_{j\ag}\rh. 
\label{9.1}
\ee 
\nit 
Under the holomorphic transformations (\ref{7}) the \Kh\, potential 
$K(\bz,z)$ itself transforms as 
\be
\del_i K(\bz,z)\, =\, F_i(z) + \bF_i(\bz), 
\label{10}
\ee 
\nit
where the holomorphic functions $F_i(z)$, $\bF_i(\bz)$ are given by 
\be 
F_i\, =\, K_{,\ag} R_i^{\ag}\, +\, i M_i, \hspace{2em} 
\bF_i\, =\, K_{,\uag} \bar{R}_i^{\uag}\, -\, i M_i. 
\label{11}
\ee 
\nit 
As is obvious from eq.(\ref{2.1}), such a change of the \Kh\, 
potential does not affect the form of the metric $g_{\ag \uag}$. 
The model defined by the lagrangean (\ref{1}) can be extended in 
several ways: by adding superpotential terms, by gauging some 
or all of the internal symmetries \ct{BW}-\ct{Mat}, by coupling 
other matter supermultiplets in linear or non-linear representations 
of the internal symmetry \ct{jw1}, or by gauging supersymmetry in 
coupling the model to supergravity \ct{Cretal}. A review discussing 
various aspects can be found in ref.\ct{Ban}. 

The presence of chiral anomalies in the internal symmetries 
(\ref{7}) restricts the usefulness of these models for 
phenomenological applications \ct{Ong,Moore,CoGo,jw3}. In 
particular, anomalies have to be removed to allow for a 
consistent gauging of the symmetries, including the chiral 
$U(1)$ symmetry (\ref{4}). To achieve this, it is necessary to couple 
additional chiral fermions in other representations of the 
symmetry group to the $\sg$-model preserving the holomorphic 
Killing vectors. To obtain chiral representations of the 
internal symmetries on the fermions whilst respecting 
supersymmetry, these fermions must be members of additional 
chiral superfields. Then the coupling amounts to embedding 
the \Kh\, manifold, on which the original superfields 
live, inside a larger \Kh\, manifold whilst preserving the 
holomorphic Killing vectors of the original model. Such a 
procedure was described in detail in ref.\ct{jw1}. 

In this letter we construct new matter representations of the 
Lie-algebra (\ref{6}), generalizing the results of \ct{jw1}.  
These new representations help to cancel the chiral anomalies 
of the $\sg$-model by allowing a more general assignment 
of charges to the matter fields. However, this gives rise to 
non-standard transformation rules of the new matter superfields 
under internal symmetries, and requires non-polynomial 
lagrangeans. A general presciption for the construction of 
these actions is given. 

In the procedure of ref.\ct{jw1} one takes chiral 
superfields transforming covariantly w.r.t.\ the diffeomorphism 
group on the \Kh\, manifold as holomorphic complex tensors. 
Introducing multi-indices $A = (\ag_1,...\ag_p)$, we write 
the components of a rank $p$ tensor as $\Psi^{\ag_1...\ag_p} 
\equiv \Psi^{A} = (N^{A}, \chi_L^{A})$. Under the special 
holomorphic diffeomorphisms $z \rightarrow z^{\prime}(z)$ 
generated by the Killing vectors $R_i^{\ag}(z)$ this yields 
the infinitesimal transformations  
\be 
\ba{lllll}
\del N^A & = & \th^i R_{i,\, B}^A N^B & \equiv & 
 \dsp{ \th^i\, \sum_{k=1}^p\, R_{i,\,\bg}^{\ag_k}(z)\, 
 N^{\ag_1 .. \bg .. \ag_p}, }\\ 
 & & \\ 
\del \hat{\chi}_L^A & = & \th^i R_{i, \, B}^A \hat{\chi}_L^B 
 & \equiv & \dsp{ \th^i\, \sum_{k=1}^p\, R_{i,\,\bg}^{\ag_k}(z)\, 
 \hat{\chi}_L^{\ag_1 .. \bg .. \ag_p}, } 
\ea 
\label{12}
\ee 
\nit
provided one takes the covariant spinor components 
\be
\hat{\chi}_L^A\, =\, \chi_L^A\, +\, \Gam_{B\bg}^{\;\;\;A} N^B 
 \psi_L^{\bg}\, \equiv\, \chi_L^{\ag_1 ... \ag_p}\, 
 +\, \sum_{k=1}^p\, \Gam_{\bg \gam}^{\;\;\;\;\ag_k} 
 N^{\ag_1 .. \gam .. \ag_p}\, \psi^{\bg}_L.
\label{13}
\ee 
\nit 
It is straighforward to check, that these fields define a 
representation of the Lie algebra (\ref{6}). Invariant lagrangeans 
for the superfields $\Psi^A$ are easily constructed by adding 
diffeomorphism invariant terms to the \Kh\, potential, the simplest 
ones being 
\be 
\Del K(\bar{\Fg},\Fg;\bar{\Psi},\Psi) = 
 g_{A\underline{A}}\, \bar{\Psi}^{\underline{A}} \Psi^A 
 \equiv g_{\ag_1\underline{\ag_1}}(\bar{\Fg},\Fg) ... 
 g_{\ag_p\underline{\ag_p}}(\bar{\Fg},\Fg)\, 
 \bar{\Psi}^{\uag_1 ... \uag_p} \Psi^{\ag_1 ... \ag_p}. 
\label{14}
\ee 
\nit 
Other possibilities involve e.g.\ contractions with curvature 
components, and terms of higher-order in $\bar{\Psi}$ and $\Psi$ \ct{jw1}. 

By construction, the transformation rules of the chiral superfields 
$\Psi^A$ under the Lie-algebra (\ref{6}) are completely fixed in terms 
of those for $\Fg^\ag$. In particular the weight under the chiral $U(1)$ 
transformations of $\Psi^A$ is fixed to be $p$ times that of 
$\Fg^{\ag}$. However, this is generally not the $U(1)$ charge 
required for the cancellation of the anomalies. 

To change this, it is sufficient to construct a scalar superfield 
$S$ with non-zero $U(1)$ charge. For the moment we take this scalar 
superfield to be dimensionless. Its $U(1)$ charge may be fixed to 
equal that of the $\sg$-model fields, i.e.\ the relative $U(1)$ 
weight is fixed to unity; for if the relative weight is not unity,
we redefine $S$ by raising it to an appropriate power. With such a 
scalar at hand, one can change the $U(1)$ weight of a tensor of 
arbitrary rank by multiplication with some power $\kg$ of $S$: 
\be 
\Psi^A\, \rightarrow\, \Psi^{(\kg)\, A}\, \equiv\, S^{\kg}\, \Psi^A.
\label{15} 
\ee 
\nit
If $q$ is the $U(1)$ charge of the scalar fields $z^{\ag}$, the 
superfield $\Psi^{(\kg)\, A}$ of rank $p$ then carries a $U(1)$ 
charge $(p + \kg) q$, with $\kg$ a free parameter. 

A scalar superfield transforming non-trivially under the Lie algebra 
(\ref{6}) can indeed be found \ct{jw1}. Writing its relevant components 
as $(s,\eta_L)$, the transformation rules are given by 
\be 
\del_i S = \lb F_i(\Fg) S\, \rightarrow\,  
\del_i s\, =\, \lb F_i(z) s, \hspace{1em} 
\del_i \eta_L\, =\, \lb \lh F_i(z) \eta_L 
 + s F_{i,\ag} \psi_L^{\ag} \rh. 
\label{16}
\ee 
\nit
These transformation rules define a consistent representation of 
the algebra (\ref{6}) for any value of the constant $\lb$, because 
the holomorphic functions $F_i(z)$ satisfy the property 
\be 
\del_i F_j\, -\, \del_j F_i\, =\, F_{j,\,\ag} R^{\ag}_i\, -\, 
 F_{i,\, \ag} R^{\ag}_j\, =\, f_{ij}^{\;\;k} F_k. 
\label{17} 
\ee 
\nit 
This follows from the definition (\ref{11}) of $F_i$, the 
Lie-algebra property (\ref{6}) of the Killing vectors, and the 
homogeneous transformation property (\ref{9}) of the Killing 
potentials $M_i(\bz,z)$. This last one holds only if the constants 
of integration in the Killing potential are properly fixed; we 
thus observe that the same choice of constants fixes the constant 
contribution to the holomorphic functions $F_i$ so as to guarantee 
the transformation rule (\ref{17}). 

Next we observe that an imaginary constant term may be present 
in the holomorphic function $F_{\th}$ for the $U(1)$ transformations 
if some of the isometries are non-linear\footnote{When the broken 
generators of the algebra all have the same charge one can show 
that this term does not vanish and the present construction 
is guaranteed.}, because such a term is present in $M_{\th}$ 
whenever $U(1)$ appears on the right-hand side of the Lie-algebra 
(\ref{6}): $F_{\th} = i a q/f^2$, with $q$ the 
elementary $U(1)$ charge of the $\sg$-model scalars and $a$ a 
dimensionless constant. On the other hand, if the $U(1)$ charge 
commutes with all other symmetries a constant term like this can 
freely be added, as a counterpart to the Fayet-Iliopoulos mechanism 
in the context of supersymmetric $\sg$-models. However it is 
generated, such a term is invisible in the transformation of $K$, 
as this only involves the real part of $F$. But it generates 
precisely the $U(1)$ transformation of $S$ we are looking for; 
indeed, taking $\lb = f^2/a$ in eqs.(\ref{16}) one gets  
\be 
\del_{\th}\, s\, =\, i q \th s, \hspace{2em} 
\del_{\th}\, \eta_L\, =\, i q \th \eta_L.
\label{18} 
\ee 
\nit 
In this way the transformations with unit relative $U(1)$ weight 
are realized on the scalar superfield $S$. 
 
Because of the special form of the transformation rule (\ref{16}) 
for $S$ and the transformation rule (\ref{10}) of the \Kh\, 
potential, the real composite superfield 
\be 
Y\, =\, \bar{S} S\, e^{-\frac{f^2}{a}\, K(\bar{\Fg},\Fg)}, 
\label{19}
\ee 
\nit 
is an invariant under the internal symmetries generated by the 
Lie-algebra of holomorphic Killing vectors and the corresponding 
transformations of $S$ and $\bar{S}$. In a different context this 
construction was used in \ct{Ber,Mar}. A physical scalar field 
of unit mass dimension and $U(1)$ weight $r$ is obtained by 
taking: $H^{(r)} = S^r/f$; the $D$-term of the real superfield 
$\Del K_H = Y^r/f^2$ defines its kinetic action. 

We can now proceed similarly with the non-standard tensor 
representations $\Psi^{(\kg)\,A}$. Multiplying the standard 
\Kh\, potential (\ref{14}) by the invariant $Y^{\kg}$ and absorbing 
the powers of $S$ in the matter superfields $\Psi^A$, one obtains
\be 
\Del K^{(\kg)}(\bar{\Fg},\Fg;\bar{\Psi}^{(\kg)},\Psi^{(\kg)})
  = e^{- \frac{\kg f^2}{a} K(\bar{\Fg},\Fg)}\, 
  g_{A\underline{A}}(\bar{\Fg},\Fg)\, \bar{\Psi}^{(\kg)\, 
  \underline{A}} \Psi^{(\kg)\,A}. 
\label{19.1}
\ee 
The transformation rule for $\Psi^{(\kg)\,A}$ follows directly 
from those of $\Psi^A$ and $S$: 
\be 
\del_i \Psi^{(\kg)\,A}\, =\, R_{i,\,B}^A(\Fg)\, \Psi^{(\kg)\,B}\, 
 +\, \frac{\kg f^2}{a}\, F_i(\Fg)\, \Psi^{(\kg)\,A}.
\label{19.2}
\ee 
These transformations are holomorphic and generally non-linear, 
although linear in the matter fields $\Psi^{(\kg)\,A}$ themselves.
Keeping this in mind, 
it is immediately obvious that the new contributions $\Del K$ 
to the \Kh\, potential are strictly invariant under these field 
transformations; hence there are no new contributions to the 
holomorphic functions $F_i(\Fg)$ in the variation of the full 
\Kh\, potential $\tilde{K} = K + \Del K^{(\kg)}$: 
\be 
\del_i \tilde{K}\, =\, F_i(\Fg)\, +\, \bF_i(\bar{\Fg}),
\label{19.3}
\ee
with the same holomorphic functions $(F_i,\bF_i)$ as for the 
original $K$ itself. This holds of course also for the special 
case of an additional scalar, with an extra term in $\tilde{K}$ 
given by $\Del K_H$ defined above. 
 
For a general theory of chiral superfields $Z^I = (\Fg^{\ag}, 
H^{(r)}, \Psi^{(\kg)\,A})$, with $H^{(r)}$ denoting any scalars 
and $\Psi^{(\kg)}$ any non-scalar matter superfields, it is now 
straightforward to compute the full Killing potentials; in terms 
of the scalar components of the superfields $z^I = 
(z^{\ag}, h^{(r)}, N^{(\kg)})$ one obtains 
\be 
\ba{lll}
\tilde{M}_i & = & 
 i \lh \tilde{K}_{,I}\, \del_i\, z^I\, -\, F_i \rh \\ 
 & & \\
 & = & M_i\, \lh 1 - \frac{rf^2}{a}\, e^{-\frac{r f^2}{a} K}\, 
  \left|h^{(r)}\right|^2 - \frac{\kg f^2}{a}\, e^{-\frac{\kg f^2}{a} K}\,  
  g_{A\underline{A}} \bar{N}^{(\kg)\,\underline{A}} 
  N^{(\kg)\,A} \rh \\ 
 & & \\
 & & +\, i e^{-\frac{\kg f^2}{a} K}\, R_{i\, \underline{A},A}\, 
  \bar{N}^{(\kg)\,\underline{A}} N^{(\kg)\,A}. 
\ea 
\label{19.4}
\ee 
For $r = \kg = 0$ the expression agrees with ref.\ct{jw1}; the last 
term can also be written in the form 
\be 
i R_{i\, \underline{A},A}\, \bar{N}^{(\kg)\,\underline{A}} 
 N^{(\kg)\,A}\, =\, \sum_{k=1}^p\, g_{\ag_1 \uag_1} ... 
 M_{i,\, \ag_k \uag_k}...g_{\ag_p \uag_p}\, 
 \bar{N}^{(\kg)\, \uag_1...\uag_p} N^{(\kg)\,\ag_1...\ag_p}. 
\label{19.5}
\ee 
Clearly, for the case of a scalar $h^{(r)}$ this term is absent. 

The algebra of symmetry transformations (\ref{7}), (\ref{19.2}) 
can now be gauged in a straightforward way by introducing 
covariant derivatives into the action, with Yukawa couplings for 
the gauginos, and the addition of a scalar potential arising from 
elimination of the auxiliary $D_i$ fields:
\be
\tilde{V}\, =\, \frac{g^2}{2}\, \left[ \tilde{M}_i (\bar{z}, z; 
\bar{h}^{(r)},h^{(r)};\bar{N}^{(\kg)},N^{(\kg)}) \right]^2.
\label{19.7}
\ee 
Being a sum of squares, the potential admits a supersymmetric 
vacuum only if all $\tilde{M}_i = 0$. For the pure supersymmetric 
$\sg$-model this is often impossible; in particular, for coset 
models $G/H$ the sum $M_i^2$ equals the curvature, a positive 
constant if $G$ is compact \ct{Aoy}. In contrast, in the presence 
of a charged matter fields, like a scalar $h^{(r)}$, it becomes
much easier: for any values of $(z,\bar{z})$ and $r/a >0$ we can 
take $\left|h\right|^2 = af^2/r\, e^{rf^2K/a}$ and $N^{(\kg)\,A} 
= 0$. This is all-right, as all  goldstone bosons $z$ now disappear 
from the physical spectrum, being replaced by massive vector boson 
states. Furthermore the vacuum expectation value of $h^{(r)}$ breaks 
the $U(1)$ symmetry, whilst supersymmetry is preserved. As a result 
a complete massive vector multiplet formed by the $U(1)$ gauge boson, 
the higgs scalar $h^{(r)}$ and a dirac fermion formed out of the 
associated gaugino and higgsino, is seen to decouple at low energy. 
If such a singlet with $r/a > 0$ is not present more complicated 
situations may arise \ct{GNH}.
\vs{1} 

\nit 
We illustrate the above construction in the context of a well-known 
model with a phenomenologically interesting particle spectrum, 
defined by the homogeneous coset space $E_6/SO(10) \times U(1)$ 
\ct{jw4,jw5}; however, our construction applies to all types of 
\Kh\, manifolds, including non-compact and non-homogeneous ones 
\ct{GS1,GS2}. The target manifold $E_6/SO(10) \times U(1)$ is 
parametrized by 16 complex fields $z_{\ag}$, $\ag = 1, ..., 16 $, 
transforming as a Weyl spinor under $SO(10)$. Their chiral fermion 
superpartners have the quantum numbers of one full generation of 
quarks and leptons, including a right-handed neutrino. The 
holomorphic Killing vectors spanning the non-linear representation 
of $E_6$ in terms of these complex fields were given in \ct{jw5}, 
whilst the extension to the case of $SO(10)$ vector and $SO(10)$ 
scalar matter superfields were presented in \ct{jw1}. We summarize 
these results here and explain how to realize the non-standard 
extensions of \Kh\, $\sg$-models presented above in this context. 

The \Kh\, potential of the model can be cast in the form \ct{jw4,jw5}
\be 
K(\bz,z)\, =\, \bz \cdot [Q^{-1} \ln (1 + Q)] \cdot z, 
\label{20}
\ee 
\nit
where the covariant $(1,1)$ tensor $Q$ is defined as 
\be 
Q_{\ag}^{\bg} = \frac{f^2}{4}\, M_{\ag \gam}^{\bg \del}\,  
  \bz^{\gam} z_{\del}, 
\label{21}
\ee 
\nit 
with $M_{\ag \gam}^{\bg \del} = 3 \del^{+ \bg}_{\; \ag} 
\del^{+ \del}_{\; \gam} - \frac{1}{2} \Gam^{+\;\;\;\;\bg}_{mn \ag} 
\Gam^{+\;\;\;\;\del}_{mn \gam}$; the matrices $\Gam^{+}_{mn}$ 
are the generators of $SO(10)$ on positive chirality spinors 
of $SO(10)$ \ct{jw5}. The dimensionful constant $f$ is the one
introduced before to assign correct physical dimensions to the 
scalar fields $(\bz,z)$. The metric derived from this \Kh\, potential 
possesses a set of holomorphic Killing vectors generating a 
non-linear representation of $E_6$: 
\be 
\del z_{\ag} = \th R^{\th}_{\ag} + \frac{1}{2}\, \og_{mn} 
 R^{mn}_{\ag} + \frac{1}{2}\, \lh \eps \cdot \bar{R}_{\ag} + 
 \bar{\eps} \cdot R_{\ag} \rh, 
\label{22}
\ee 
\nit 
with 
\be 
\ba{ll}
\dsp{ R^{\th}_{\ag} = \frac{i}{2}\, \sqrt{3} z_{\ag}, } & \dsp{ 
 R^{mn}_{\ag} = - \frac{1}{2}\, \lh \Gam^+_{mn} \cdot z\rh_{\ag}, 
 }\\
 & \\
\dsp{ \bar{R}^{\bg}_{\ag}\, =\, \frac{i}{f} \del^{\bg}_{\ag}, } & 
\dsp{ R_{\bg \ag} = - \frac{if}{4}\, M_{\bg \ag}^{\gam \del}\, 
 z_{\gam} z_{\del}. }
\ea 
\label{23}
\ee 
\nit 
The corresponding Killing potentials are 
\be 
\ba{ll} 
\dsp{ M^{\th} = \frac{1}{f^2 \sqrt{3}}\, -\, \frac{1}{2}\, \sqrt{3}\, 
 \bz^{\ag} K_{, \ag}, } & \dsp{ M^{mn} = - \frac{i}{2}\, \bz^{\ag}   
 \Gam^{+\;\;\;\bg}_{mn \ag} K_{,\bg}, } \\
& \\
\dsp{ \bar{M}^{\bg} = - \frac{1}{f}\, K_,^{\bg}, } & \dsp{ M_{\bg} = 
 - \frac{1}{f}\, K_{, \bg}. } 
\ea 
\label{24}
\ee 
\nit 
Observe the presence of the constant term in the $U(1)$ Killing 
potential $M^{\th}$ which is required to close the Lie algebra on 
the Killing potentials. 

Having in hand the \Kh\, and Killing potentials, we can construct 
the holomorphic functions $F_i(z)$ from eq.(\ref{11}). A 
straightforward calculation yields 
\be 
\ba{ll} 
\dsp{ F^{\th} = \frac{i}{f^2 \sqrt{3}}, } & F^{mn} = 0, \\
\bar{F}^{\bg} = 0, & \dsp{ F_{\bg} = - \frac{i}{f}\, z_{\bg}. }
\ea  
\label{25}  
\ee 
\nit 
For the variation of the \Kh\, potential this gives the standard 
result \ct{jw4}
\be 
\del K\, =\, \frac{i}{f}\, \lh \eps \cdot \bz - \bar{\eps} \cdot 
 z \rh. 
\label{25.1}
\ee
\nit
Next we insert the expressions for the $F_i(z)$ into eq.(\ref{16}) 
to get the superfield transformations 
\be 
\del S\, =\, \lb \lh \frac{i}{f^2 \sqrt{3}} \th - \frac{i}{f} 
     \bar{\eps} \cdot \Fg \rh\, S. 
\label{26}
\ee 
\nit 
Taking $\lb = 3 f^2/2$ gives the desired result of relative unit 
$U(1)$ weight: 
\be 
\del S\, =\, \frac{i}{2}\, \sqrt{3}\, \th S\, -\, 
 \frac{3i}{2}\, f\, \bar{\eps} \cdot \Fg\, S,
\label{27} 
\ee 
\nit
which has the same $U(1)$ weight as the $\underline{16}$ of 
goldstone scalars $z^{\ag}$ defining the unit of $U(1)$ charge. 
This result is in agreement\footnote{We take the opportunity 
of pointing out a misprint in eqs.(6.6) in \ct{jw1}, where a minus 
sign is missing in the last line.} with previous results \ct{jw1}. 

The compensating superfield $S$ can now be used to construct other 
non-standard representations of the $E_6$ algebra. For example, 
a scalar $H$ and a vector $\Psi_m$ complete the set of complex chiral 
superfields to form a $\underline{27}$ of $E_6$, provided the $U(1)$ 
charges are assigned correctly; indeed, its decomposition under 
$SO(10)$ reads 
\be 
\underline{27}\, \rightarrow\, \underline{16}(1) + 
 \underline{10}(-2) + \underline{1}(4), 
\label{27.1}
\ee
where the numbers in parentheses denote the relative $U(1)$ 
weights. With this choice of matter content, the cancellation of 
chiral anomalies of the full $E_6$ isometry group is achieved. 

An $SO(10)$ scalar with relative $U(1)$ weight $r$ is now obtained by
taking $H^{(r)} = S^r/f$; its transformations under the full $E_6$ are  
\be
\del H^{(r)} = i\frac r2 \sqrt{3}\, \th H^{(r)}\, -\, i\frac {3r}2 f\,
 \bar{\eps} 
\cdot \Fg\, H^{(r)}.
\label{27.2} 
\ee 
A superfield $\Psi_m$, $m = 1,...,10$, in the vector representation 
of $SO(10)$, is an irreducible part of a bi-spinor:  
\be 
\Psi_m\, =\, - \frac{1}{16}\, \mbox{Tr}(\tilde{\Psi} C \bar{\Sg}_m), 
\label{28}
\ee 
\nit 
where $\tilde{\Psi}_{\ag\bg}$ is a bi-spinor, $(\Sg_m, \bar{\Sg}_m)$ 
are the 10-dimensional analogues of the Pauli-matrices and $C$ the 
corresponding charge conjugation matrix \ct{jw5}. The standard 
representation of $E_6$ on the bi-spinor is obtained by the same 
reduction of the direct product of the representations defined by the 
Killing vectors (\ref{23}). This gives 
\be 
\del \Psi_m\, =\, i \th \sqrt{3}\, \Psi_m\, -\, \og_{mn} \Psi_n\, 
 -\, if \bar{\eps} \cdot \lh \Gam^+_{mn} + 3\, \del^+_{mn} \rh \cdot 
 \Fg\, \Psi_n.
\label{29}
\ee 
\nit 
To construct a non-standard representation we consider the rescaled 
chiral superfield $\Psi_m^{(\kg)} = S^{\kg} \Psi_m$. Its 
transformations can be worked out in a straightforward way, 
with the result 
\be
\del \Psi_m^{(\kg)}\, =\, \frac{i}{2}\, \sqrt{3} \lh \kg + 2 \rh \th\, 
 \Psi_m^{(\kg)} - \og_{mn} \Psi_n^{(\kg)} - if \bar{\eps} \cdot 
 \lh \Gam^+_{mn} + \frac{3}{2}\, (\kg + 2)\, \del^+_{mn} \rh \cdot 
 \Fg\, \Psi_n^{(\kg)}. 
\label{30}
\ee 
\nit 
For $\kg = -4$ the relative $U(1)$ charge of the 10-vector 
$\Psi_m^{(-4)}$ becomes $-2$, as required by the anomaly 
cancellation. Note also, that $\kg = -2$ would give an $SO(10)$ 
vector of zero charge. In the following we take the 
values $r = 4$ and $\kg = -4$ for the relative $U(1)$ weights of
$H$ and $\Psi_m$ understood, and drop the superscripts.  

An invariant kinetic action for these matter superfields is 
obtained as an application of eqs.\ (\ref{19}) 
and (\ref{19.1}):
\be 
\Del K\, =\, |H|^2 e^{-6f^2 K(\bar{\Fg},\Fg)}\, 
 +\, g_{mn}(\bar{\Fg},\Fg)\, \bar{\Psi}_m \Psi_n\, 
 e^{6 f^2 K(\bar{\Fg},\Fg)}. 
\label{31}
\ee 
\nit
Here the induced metric for the 10-vector representation is 
\be 
g_{mn}(\bar{\Fg},\Fg)\, =\, -\, \frac{1}{16}\, 
 g^{\;\bg}_{\ag}(\bar{\Fg},\Fg)\, g^{\;\del}_{\gam}(\bar{\Fg},\Fg)\, 
 \lh C \bar{\Sg}_m \rh^{\ag \gam}\, \lh \Sg_n C \rh_{\bg \del}. 
\label{32}
\ee 
The factor 1/16 is introduced to normalize the metric to $\del_{mn}$ 
in the limit $\Fg = 0$. It remains to construct the extended Killing 
potentials. Using the general results (\ref{19.4}), (\ref{19.5}) we 
find 
\be 
\ba{lll}
\tilde{M}_i & = & \dsp{ M_i\, \lh 1 - 6 f^2 e^{-6f^2 K} 
  |h|^2 + 6 f^2 e^{6f^2 K} g_{mn} \bar{N}_m N_n \rh }\\
 & & \\
 & & \dsp{ -\, 
   \frac{1}{8}\, e^{6f^2 K} M_{i, \ag}^{\;\;\bg}\, g_{\gam}^{\;\del}\, 
   \lh C\bar{\Sg}_m \rh^{\ag\gam} \lh \Sg_nC \rh_{\bg\del}  
   \bar{N}_m N_n, }
\ea 
\label{33}
\ee 
with the Killing potentials $M_i$ given by the expressions (\ref{24}). 
In the case the full $E_6$ symmetry is gauged, the scalar potential 
is given by the implementation of eq.(\ref{19.7}). However, in that 
case all the goldstone bosons disappear from the spectrum as a result 
of the Brout-Englert-Higgs effect; in the unitary gauge $z_{\ag} = 0$
the potential then simplifies to the form 
\be 
\ba{lll}
\tilde{V}_{unitary} & = & {\dsp \frac{g^2}{2}\, \sum_i\, \left[ 
 \tilde{M}_i (\bar{h},h;\bar{N}_m,N_m) \right]^2 }\\ 
 & & \\
 & = & \dsp{ \frac{g^2}{2}\, \lh \frac{1}{f^2 \sqrt{3}} - 2 \sqrt{3}\, 
 |h|^2 + \sqrt{3}\, |N_m|^2 \rh^2 } \\
 & & \\ 
 & & \dsp{ +\, \frac{g^2}{2}\, |\bar{N}_m N_n - 
 \bar{N}_n N_m|^2. }
\ea 
\label{34} 
\ee 
The potential possesses a large continuous set of supersymmetric 
minima given by the solutions {\em Im}$(N_m) = 0$, and 
\be
|h|^2\, =\, \frac{1}{6f^2}\, +\, \frac{1}{2}\, |N_m|^2. 
\label{35}
\ee 
It follows that $|h| \neq 0$ and the $U(1)$ gauge symmetry is always 
broken; a solution with $|N_m| = 0$ is possible, preserving 
$SO(10)$. However, solutions with $|N_m| \neq 0$ breaking $SO(10)$ 
are allowed, and expected in a next stage of symmetry breaking. 

Clearly the above procedure is quite powerful for constructing 
consistent effective actions for supersymmetric field theories. 
Interesting applications also include the Grassmannian 
coset manifolds, describing hierarchies of $SU(N)$ symmetries 
\ct{BW}-\ct{jw2},\ct{Ban}. Furthermore our construction can be 
extended to supergravity, allowing a larger class of supersymmetric 
unification models to be studied, e.g.\ based on 
the coset $E_6/SO(10) \times U(1)$ discussed here in Minkowski 
space-time. A more detailed analysis of scalar potential minima, 
mass spectra and residual symmetries is called for. We intend 
to elaborate on these issues in a separate paper \ct{GNH}.

\end{document}